\documentclass[12pt,epsf]{article}

\usepackage{latexsym}
\usepackage{epsfig}
\usepackage{amsfonts}

\begin{document}

\begin{center}
{\bfseries  \Large Monte-Carlo calculation of the lateral Casimir
forces between rectangular gratings within the formalism of
lattice quantum field theory} \vskip 5mm

Oleg Pavlovsky,$^{\dag}$  Maxim Ulybyshev$^{\ddag}$

\vskip 5mm

{\small {\it Institute for Theoretical Problems of Microphysics, \\
Moscow State University}
\\
$\dag$ {\it E-mail: ovp@goa.bog.msu.ru } $\ddag$ {\it E-mail:
ulybyshev@goa.bog.msu.ru }}
\end{center}

\vskip 5mm

\begin{center}
\begin{minipage}{150mm}
\centerline{\bf Abstract} We propose a new Monte-Carlo method for
calculation of the Casimir forces. Our method is based on the
formalism of noncompact lattice quantum electrodynamics. This
approach has been tested in the simplest case of two ideal
conducting planes. After this the method has been applied to the
calculation of the lateral Casimir forces between two ideal
conducting rectangular gratings. We compare our calculations with
the results of PFA and ``Optimal'' PFA methods.
\end{minipage}
\end{center}

\vskip 3mm


\vskip 3mm


\vskip 10mm

%%%%%%%%%%%%%%%%%%%%%%%%%%%%%%%%%%%%%%%%%%%%%%%%%%%%%%%%%%%%%%%%%%%%%%%%%%%%%%%

\section*{Introduction}
The Casimir effect attracts significant attention in the last few
years due to its important role in micro- and nano- mechanics
\cite{book}. There are two main problems in this area: calculation
of Casimir forces for bodies that are interesting from
experimental point of view; consideration of both the complicated
shape and electromagnetic properties of the interacting bodies. It
is also important to take into account temperature effects.
Radiative corrections are small in QED \cite{bordag}. But they may
play more important role in case of non-abelian fields. Our
approach makes it possible to take them into account.

There are many numerical methods have been proposed for
calculation of the Casimir energy. Let us discuss some of them.

One of the most efficient methods for Casimir energy calculation
was derived in \cite{mit1}. In this case the stress tensor and the
net force on a body is obtained from Euclidean Green's function.
It can be calculated numerically by using of standard
electromagnetic methods (for example, by Finite Elements Method or
Boundary Elements Method).

Another interesting numerical method for the Casimir energy
calculation was proposed in \cite{wline}. This worldline algorithm
is formulated for calculations of Casimir forces induced by
scalar-field fluctuations with Dirichlet boundary conditions for
various geometries. Unfortunately this method was not extended to
Neumann boundary conditions.

Our approach is based on Quantum Field Theory on the lattice. It
is one of the common frameworks in modern science. It is based on
the Monte-Carlo calculations and provides an efficient tool to
study all the above-mentioned problems. We have already studied
the Casimir interaction of the Chern-Simons surfaces by using
lattice QED \cite{ijmpa,ichaya}. In the present paper we consider
Monte-Carlo calculation of the Casimir interaction between ideal
conductors and between dielectric bodies. As an example, we study
the Casimir effect for two rectangular gratings. The normal part
of the vacuum force in this system has already been calculated
\cite{mar}. But there exist also lateral forces, which are of
significant experimental and technical interest.

\section{Ideal conductors and dielectric bodies in QED on the lattice}
\subsection{Conductor boundary conditions}

In this paper we use the 4-dimensional hypercubical lattice and
the simplest action of noncompact lattice QED in Euclidean time:

\begin{equation}
 S=\frac{\beta}{2} \sum_x \sum_{\mu < \nu}
\theta^2_{\mu\nu}(x), \label{actionL}
\end{equation}
where the link and plaquette variables are defined as follows:
\begin{equation}
U_{\mu}(x)=e\, a\, A_{\mu}, \label{link}
\end{equation}
$$\theta_{\mu \nu}(x)=\triangle_\mu U_{\nu}(x)-\triangle_\nu U_{\mu}(x),$$
$$\triangle_\mu U_{\nu}(x)=U_{\nu}(x+\hat{\mu}) - U_{\nu}(x),$$
where $a$ is the lattice step, $\beta=1/e^2$. The lattice is
formed by the sites. They are vertices of 4-dimensional cubes. The
links are edges of these cubes and plaquettes are their sides. The
functional integrals for physical quantities are calculated by
means of the Monte-Carlo method. It means that we generate field
configurations (sets of link variables) with statistical weight
$e^{-S}$.  After it the functional integrals are calculated as
field configurations averages.

Our first task is formulation of the ideal conductor boundary
condition on the lattice. Both electric and magnetic fields are
pushed away from the interior of conductor. So the boundary
conditions for the fields can be written as:
\begin{eqnarray}
&&E_\parallel \vert_{S}=0 ,\qquad H_n \vert_{S}=0
.\label{condition1}
\end{eqnarray}
Because the hypercubical lattice is used in our approach, we will
approximate a surface by a set of plane polygons. Each polygon
consists of plaquettes.

Let us consider a surface that consists of one polygon (normal to
the z-axis). More complicated surfaces can be treated similarly.
Expressions (\ref{condition1}) are the boundary conditions for the
field strength tensor $F_{\mu\nu}$.  Due to the connection between
lattice variables and fields (\ref{link}), the boundary conditions
can be reformulated in terms of plaquette variables:

\begin{equation}
\theta_{41}(x)=0, \quad \theta_{42}(x)=0, \quad
 \theta_{12}(x)=0 \ .
 \label{bound-cond}
\end{equation}

All plaquettes in these conditions form 3-dimensional sublattice
that describes the conductor surface. Pure gauge is the only type
of field configuration that satisfies this conditions. It means
that any link variable inside this sublattice can be written as:
\begin{equation}
U_{i}=\alpha(x+ \hat i)-\alpha(x),
 \label{gauge}
\end{equation}
where $\alpha(x)$ is an arbitrary function defined on the sites of
3-dimensional sublattice. This scheme is valid for any surface.

The functional integrals should be rewritten, taking into account
the conductor boundary conditions. Our functional integrals must
be formulated in terms of independent field variables $A_\mu (x)$,
but in our case some of them are not independent due to the
conditions (\ref{condition1}). This problem can be solved by
choosing the function $\alpha(x)$ as an independent variable which
parameterizes the fields on the boundary surface.

So the functional integral for the partition function takes the
form:
$$
Z=\int D{\tilde A}_\mu D\alpha e^{(iS[{\tilde A}_\mu, \alpha])}\ ,
$$
where $\tilde A$ is the vector potential of the fields outside the
boundary surface.

Physical quantities are calculated by means of the Euclidean
functional integral:
\begin{equation}
\langle F\rangle={\int D{\tilde A}_\mu D\alpha F[A_\mu]
e^{-S_{eucl.}[{\tilde A}_\mu, \alpha]} \over \int D{\tilde A}_\mu
D\alpha e^{-S_{eucl.}[{\tilde A}_\mu, \alpha]} }.
\label{average_gen}
\end{equation}
It is very important to note also that quantity $F[A_\mu]$ is
gauge invariant.

Now let us consider lattice discretization of(\ref{average_gen}).
The Euclidean action $S_{eucl.}[{\tilde A}_\mu, \alpha]$ and the
observable $F[A_\mu]$ can be rewritten in terms of lattice
variables: $S_{Eucl. Lat.}$ and $F[U_\mu]$. Then the functional
integral is Boltzmann-type field configuration average with
statistical weight $\exp (-S)$. The lattice field configurations
consist of the link variables $ U_\mu $ and the site variables
$\alpha (x)$. The link variables describe the electro-magnetic
field outside the boundary surface. The site variables are defined
on the lattice sites that belongs to boundary surface.  The
function $\alpha (x)$ parameterizes the pure gauge fields on this
surface.

 The pure gauge field on the
surface can be transformed into the zero-value fields
$\alpha(x)=0$ by means of a suitable lattice gauge transformation.
After such transformation, the action $S_{Eucl. Lat.}$ and the
observable $F$ do not change their values. It means that in order
to implement the ideal conductor boundary conditions it is
sufficient to generate field configurations with zero-value link
variables on the boundary surface.

Let us consider the following example: the Casimir interaction of
two parallel conducting planes.

\begin{figure}[t]
 \begin{center} \epsfysize=60mm \epsfbox{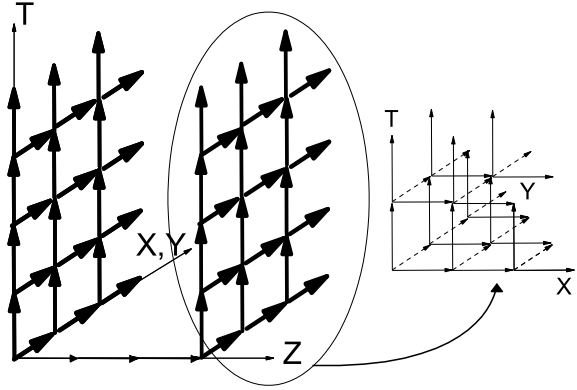}
 \end{center}\caption{
Two parallel conducting planes: sublattices of links with zero
value link variables.}
\end{figure}

The boundary links are always taken equal to zero during the
generation procedure. In Fig.~1, these boundary links are
bold-marked. The boundary links form two 3-dimensional sublattices
at $z=0$ and $z=N$, where $N$ is lattice size. The 3-dimensional
structure of these sublattices is shown in Fig.~1 as well. These
boundary conditions can be written explicitly as
$$
U_{i}(x,y,z,t)=0 \quad \hbox{if} \quad i=1,2,4 \quad \hbox{and}
\quad z=0,N \ .
$$
The periodic boundary conditions are used for the other three
coordinates ($x$, $y$ and $t$).

This approach can be used to implement the ideal conductor
boundary conditions for any surface shape.

\subsection{Description of dielectric bodies in QED on the lattice}

The simplest dielectric (we neglect anisotropy and dielectric
constant frequency dependence) can be described by the following
action:
$$
S= {1 \over 4} \int\limits_{\bar V} F_{\mu\nu} F^{\mu\nu} dV + { 1
\over 2} \int\limits_{V}( \varepsilon \sum\limits_{i=1}^{3} F_{0i}
F^{0i} + \sum\limits_{i<j} F_{ij}F^{ij}) dV.
$$
Here $V$ is the 4-dimensional volume occupied by dielectric body
and $\varepsilon$ is the static dielectric constant. This action
can be rewritten in terms of lattice quantities:
$$
 S=\frac{\beta}{2} \left( \sum_{x \in \bar V} \sum_{\mu < \nu}
\theta^2_{\mu\nu}(x)+ \sum_{x \in V} \left( \varepsilon \sum\limits_{i=1}^{3} \theta^2_{0i}(x) +  \sum\limits_{i<j}\theta^2_{ij}(x) \right)  \right).
$$
We shall use it for field configurations generation. The
dielectric constant frequency dependence is very important for
real materials and has to be taken into account. Firstly, we fix
the gauge:
$$
A_4=0 \ .
$$
Then we make the Fourier transform:
$$
A_i={1 \over \sqrt{2\pi} } \int \tilde A_i e^{-i\omega \tau} d
\omega, \quad i=1,2,3.
$$
Here $\omega$ is the imaginary frequency, since we make the
Fourier transform in Euclidean time. After this, the action
becomes ``block-diagonal'':
$$
S_{\mbox{\small eucl}}={1 \over 2 } \int d\vec r d\omega \left(
{\tilde F_{12}}^2 + {\tilde F_{13}}^2 + {\tilde F_{23}}^2 +
\varepsilon(i \omega) \omega ^2 (\tilde A_{1}^2 +  \tilde A_{2}^2
+ \tilde A_{3}^2) \right).
$$
So we can generate field configurations for each $\omega$
independently and take into account the $\varepsilon(\omega)$
dependence.

\section{The lattice observable for the ground state energy}

The corresponding lattice observable should give (after the
Monte-Carlo averaging) the ground state energy of a system:
\begin{equation}
\int_V \langle 0 | T^{00} | 0 \rangle d \vec x \ .
 \label{int_energy}
\end{equation}
Our task now is to define such lattice observable.

Lattice system can be treated as a certain $N$-dimensional
quantum-mechanical system where the number of degrees of freedom
$N$ tends to infinity. For simplicity we will firstly consider the
one-dimensional problem of quantum particle moving in potential
$V(x)$. The lattice formulation of this task: the path in
functional integral is a set of the coordinate values at different
moments of time ($x(t_i)=x_i,$ $i=0...N$). The boundary conditions
in the Euclidean time direction are taken periodic: $x_0=x_N$.
Such a path is an analog of the Euclidean field configuration in
Lattice Field Theory. The Monte-Carlo method for functional
integral calculation is based on generation of a set of such
lattice configurations with statistical weight proportional to
$\exp(^{-S_{eucl}[x] / \hbar})$. The Euclidean action is
$$
S_{\mbox{\small eucl}}=a \sum_{n=1}^N \left( { m {(x_n-x_{n-1})}^2
\over 2a^2} +V(x_n) \right),
$$
where $a$ is the time step. It is well known \cite{creutz} that
any coordinate $x_n$ is distributed with the probability density
of the vacuum state. It means that if one calculates the
configuration average of a certain lattice observable (for
example, the potential energy), one gets the vacuum average of
this observable (the average potential energy):

$$
\langle 0 | V | 0 \rangle=\langle V(x_n)\rangle={1 \over
N_{conf}}\sum\limits_{conf} V(x_n).
$$

In order to obtain the full energy of the system, we need to
calculate the average of the kinetic energy as well. This problem
was studied in  \cite{Ceperley}, the corresponding lattice
observable for the kinetic energy is given by
\begin{equation} \langle 0 | T
| 0 \rangle=\langle -{m\over2}{(x_{n+1}-x_n)^2\over a^2
}+{\hbar\over2a}\rangle.\label{T_direct_lim}
\end{equation}

The naive expression for the kinetic energy lattice observable
(the first part in the sum (\ref{T_direct_lim})) diverges in the
continuous limit due to fractal structure of the trajectories.
This divergence is directly cancelled in (\ref{T_direct_lim}).
This procedure can be performed for any lattice models (and for
lattice field theory too), and thus we have the method for direct
calculation of the full energy on the lattice.

Let us consider noncompact lattice QED with the Hamiltonian
density:
\begin{equation}
{\cal H} = {1 \over 2} \left( \sum\limits_{i=1}^3 {\pi_i(\vec
x)}^2 + \sum\limits_{i<j} {F_{ij}(\vec x)}^2 \right), \label{ham}
\end{equation}
where the ``field momentum'' $\pi_i=F_{0i}$ is the conjugated
quantity to the field $A_i$. The vacuum expectation value of the
second part of the expression (\ref{ham}) can be calculated
directly by the field configuration averaging. The averaging of
the first part of (\ref{ham}) is performed by the same way as the
kinetic part of one-dimensional quantum mechanical system: we
calculate the same observable, but with opposite sign at
${\pi_i(\vec x)}^2$. The final expression for vacuum expectation
value of the full energy reads

\begin{equation}
\langle 0 | H | 0 \rangle =\langle {\beta \over 2} \left(
\sum\limits_{x, i}\left( -\theta^2_{4i}(x) \right) +
\sum\limits_{x, i<j} \theta^2_{ij}(x) \right) \rangle.
\label{observ}
\end{equation}

The renormalization procedure for the lattice observable
(\ref{observ}) has three parts. Firstly, the singular contribution
in (\ref{T_direct_lim}) that corresponds to fractal structure of
the trajectories has to be eliminated. So we calculate the energy
density $\langle 0 | T^{00}(\vec x) | 0 \rangle$ on the free
lattice and then subtract it from the energy density calculated in
presence of the boundaries.
$$
 \langle 0 | T^{00}(\vec x) | 0 \rangle_{\mbox{\small bound}}-\langle 0 | T^{00}(\vec x) | 0 \rangle_{\mbox{ \small {free
 lattice}}}.
$$
As a result, one gets the difference between the Hamiltonian
densities with and without interacting bodies. It is important to
note that $\langle 0 | T^{00}(\vec x) | 0 \rangle_{\mbox{\small
{free lattice}}}$ is just a constant.

But this is not the end of the story. Any bodies have the Casimir
self-energies which are connected with their boundaries. The
Casimir self-energy calculation is an important task, e.g., for
the dynamical Casimir effect, but in our case this type of the
Casimir energy should be eliminated because we are interested only
in interaction between bodies.

Finally, we have the following three stages of the renormalization
procedure:

\begin{enumerate}

\item Calculation of the energy density  $\langle 0 |
T^{00}(\vec x) | 0 \rangle$ in presence of the boundary conditions
(interacting bodies) and subtraction of the ``fractal divergence''
constant  $\langle 0 | T^{00}(\vec x) | 0 \rangle_{\mbox{\small{
free lattice}}}$.

\item Calculation of the energy density $\langle 0 | T^{00}(\vec
x) | 0 \rangle_{\mbox{\small {self-energy}}}$ for all interacting
bodies individually.

\item Subtraction of this Casimir self-energy density $\langle 0
| T^{00}(\vec x) | 0 \rangle_{\mbox{\small {self-energy}}}$ from
the Casimir energy density of the interacting bodies after
elimination of the ``fractal  divergence''. As a result, one
obtains the renormalized Casimir energy densuty for the system of
the interacting bodies.

\end{enumerate}

\section{Results of the numerical calculations}

We use the standard ''heat bath'' method \cite{makeenko} for field
configurations generation. Also we use the anisotropic lattice to
improve the accuracy of our calculations. The anisotropic lattice
formalism is similar to the one applied in \cite{deformed} to
study the finite temperature effects, but in the present work we
deform $z$-direction instead of $t$-direction.

All link variables are connected with the field components in the
usual way:
$$U_{\mu}=e\, a\, A_{\mu}, \mu=1,2,4
$$
except links in $z$-direction:
$$U_{3}=\alpha e\, a\, A_{3}.
$$
Here $a$ is the lattice step in all directions except $z$; $\alpha
a$ is the lattice step in $z$-direction. The action can be written
as
\begin{equation}
S=\frac{\beta}{2} \Big\{ \alpha \sum_x \sum_{\mu < \nu; \mu, \nu
\neq 3} \theta^2_{\mu\nu}(x) +\frac{1}{\alpha} \sum_x \sum_{\nu}
\theta^2_{3 \nu}(x) \Big\}. \label{action_dif}
\end{equation}

The continuous limit for the new observable is different from the
traditional methods used, for example, in lattice QCD. It is well
known that in the continuous theory the Casimir energy (without
radiative corrections) is independent of the electron charge. This
property is exactly reproduced in noncompact lattice
electrodynamics for the following reasons. Firstly, the phase
transition is absent in this theory, thus  $\beta$ in
(\ref{action_dif}) plays only the role of parameter that
determines the numerical values of link variables. Therefore, if
one calculates the Casimir energy in noncompact electrodynamics
without fermions, the value of $\beta$ can be chosen arbitrarily
within sufficiently broad range and is defined solely by the
calculation convenience.

This independence of results from $\beta$ additionally manifests
itself in the fact that Casimir calculations do not actually
require any knowledge about the physical volume of the lattice and
the value of lattice step $a$. We obtain the dimensionless energy
in ${1 \over a}$ units.  All geometrical sizes of the considered
bodies are defined in the lattice step units. Therefore, the
physical value of the lattice step $a$ fixes all sizes and real
physical value of the Casimir energy.

In noncompact electrodynamics, the rotational symmetry is restored
automatically at sufficiently large lattice distances (this was
shown, e.g., in \cite{rotational}). Therefore, the continuous
limit is only the transition $N\rightarrow \infty$. In other
words, we should find that the correct dependence of the Casimir
energy on the distance between the bodies is restored at large
lattice distances.

This difference between noncompact QED and non-abelian theories
can be understood from the following simple arguments. We should
decrease the lattice step relatively to a typical scale of the
theory to achieve the continuous limit. The Compton wavelength of
 glueball is such a scale in non-abelian theories. We vary the
relation of the lattice step  to this characteristic length by
changing $\beta$. So we can make the lattice step small in
comparison with the glueball scale. It means also that the
correlation length tends to infinity in the lattice units. The
situation is quite different in non-compact Abelian theory. There
are no glueballs in noncompact Abelian theory without fermions in
the continuous limit. The correlation length is just a lattice
artifact of discretization in this theory. The scale in this
theory is defined by the sizes of the interacting bodies. It means
that the continuous limit in this theory is archived in the case
when the sizes of the bodies (in the lattice units) become large
and all the lattice sizes tend to infinity.

\begin{figure}[h]
 \begin{center} \epsfbox{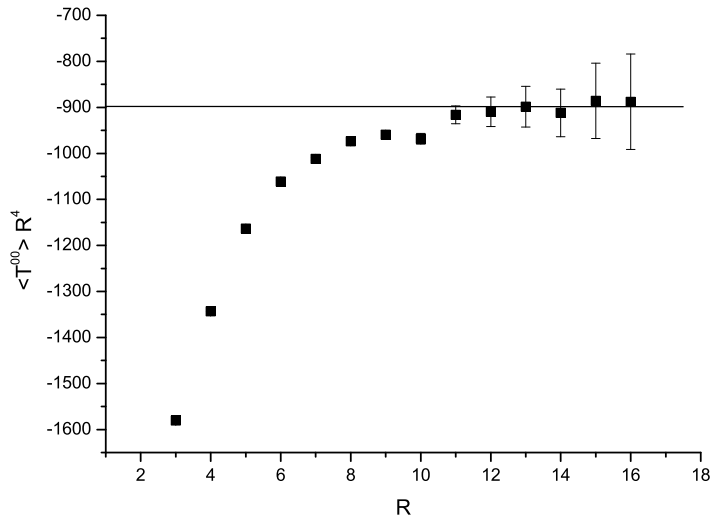}
 \end{center}\caption{ Calculation of $\langle 0 | T^{00}(\vec x) | 0
\rangle  R^4$ in case of the periodic boundary conditions, $R$ is
the size of the space region in the lattice step units. Lattice
sizes: $64 \times 64 \times 64 \times R$. The line is the
analytical answer  \cite{mamaev}.}
\end{figure}

The restoration of the continuous limit in the Casimir problem on
the lattice is illustrated in Fig.~2. In this calculation the
Casimir energy of the electro-magnetic field is studied in the
case of the space-periodic boundary conditions. This problem
provides a good test for our method because there is the
analytical solution in this case \cite{mamaev}. The lattice
observable is the vacuum energy density calculated by using the
renormalization procedure which we discussed above.

The energy density of the electro-magnetic field in case of the
periodic boundary conditions is a constant \cite{mamaev} and is
proportional to ${1 / R^4}$, where $R$ is the size of the space
region where we confine the fields by the periodic boundary
conditions (this statement is obvious from dimensional reasons).
Therefore, the quantity  $\langle 0 | T^{00}(\vec x) | 0 \rangle
R^4$  should be constant for large size of the lattice, when the
continuous limit is reproduced.

In Fig. 2 one can see that our lattice calculation of the energy
density tends to the analytical result at large lattice distances
$R$. If the lattice distance $R$ is about 11-12 lattice units, the
lattice results are very close to the analytical ones, and we can
conclude that the continuous limit is achieved just at these
lattice distances.

We test our method on the well-known problem of the Casimir
interaction of two parallel conducting planes. In Fig. 3 the
lattice calculation of the vacuum energy of the electro-magnetic
field is presented for this model (we study the planes with quite
large size due to the continuous limit reasons). We consider two
parallel planes which are separated by the distance $D$ in
$z$-direction.  The calculations are performed on the anisotropic
lattice with the following sizes: $32 a$ at $x$, $y$ and $t$
directions, $R a_z$ at $z$ direction, ${a_z / a }={1 / 3}$. The
analytical result for this problem is well known ($c=\hbar=1$):
$$
E = {\pi^2 S \over { 720 D^3}},
$$
where $S$ is the area of the planes, $D$ is the distance between
the planes. We fit the calculated points by the function $E = {P_1
/ R^3}$ using the Least Squares method. As one can find, the
dependence of the Casimir energy on the distance between planes is
reproduced correctly by our lattice method. In order to compare
this lattice result with the analytical answer, one should note
that all geometrical sizes are proportional to the lattice step
$a$, while the energy is proportional to $a^{-1}$. This means that
\begin{equation}
 E ={\pi^2 N^2 \over { 720 \alpha^3}} {1 \over R^3} {1 \over
a}. \label{anal_casim}
\end{equation}
The lattice steps in $x$, $y$ and $t$ directions are equal to $a$,
but the lattice step is equal to  $\alpha a$ in deformed $z$
direction. The area of the planes is equal to $S=(Na)^2$, where
$N$ is the lattice size in $x$ and $y$ directions; the distance
between planes is $D=\alpha a R$.

\begin{figure}[h]
\begin{center}
 \epsfbox{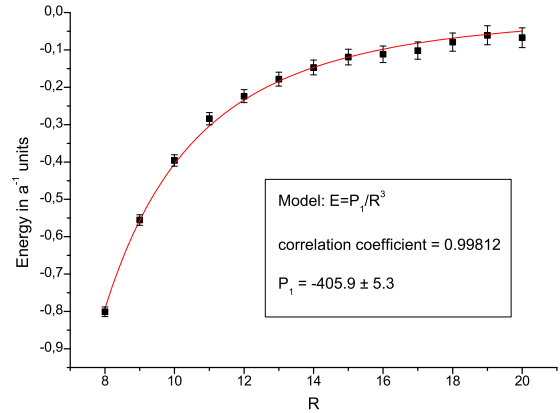} \end{center} \caption{Calculation of
 the Casimir interaction energy for two parallel conducting planes. $R$ is the distance between planes
 in the lattice steps: $D=\alpha a R$. The fitting line is found by the Least Squares
 method: $y={ P_1 \over R^3}$. All calculation are performed on
 the anisotropic lattice with the deformation coefficient ${a_z / a_{xyt} }={1 / 3}$. The lattice sizes:
 $32 a \times 32 a \times 32 a \times R a_z$. }
\end{figure}

So we have to compare the dimensionless coefficient at ${1 \over
R^3}$ in the expression for the Casimir energy (\ref{anal_casim})
with the coefficient $P_1$ calculated by fitting the numerical
data using the function $E = {P_1 / R^3}$. The analytical answer
for this coefficient is $P^{\small analyt}_1=379$ (the parameters
of the lattice are shown in Fig. 3). In this test problem one can
find that the numerical accuracy of our approach is about $5 \%$.

Let us apply our method to the calculation of the Casimir forces
between two conducting surfaces with complicated shapes. We
consider two rectangular gratings (see  Fig. 4).

\begin{figure}[t]
\begin{center}
 \epsfbox{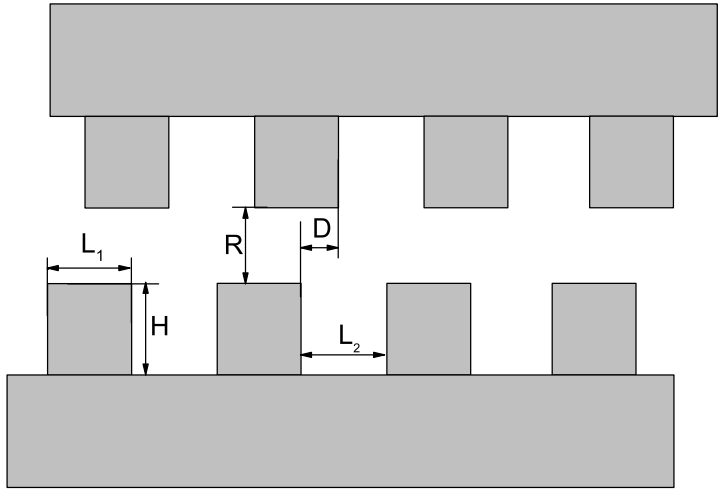} \end{center} \caption{Geometrical parameters of the gratings.}
\end{figure}

We perform the calculation for the ideal conducting gratings with
the following parameters: the width of the ``tooth'' and the
distance between them are $L_1=L_2=7 a$; the height of the
``tooth'' is $H=7 a_z$; the distance between the gratings (between
the tops  of the ``teeth'') is $R=8 a_z$. All sizes are chosen
sufficiently large to restore the continuous limit. The lattice is
the same that in the previous calculation for two planes: ${a_z /
a}= {1 / 3}$.

The result of the Casimir energy calculation  for the two gratings
is shown in Fig. 5.

\begin{figure}[h]
\begin{center}
 \epsfbox{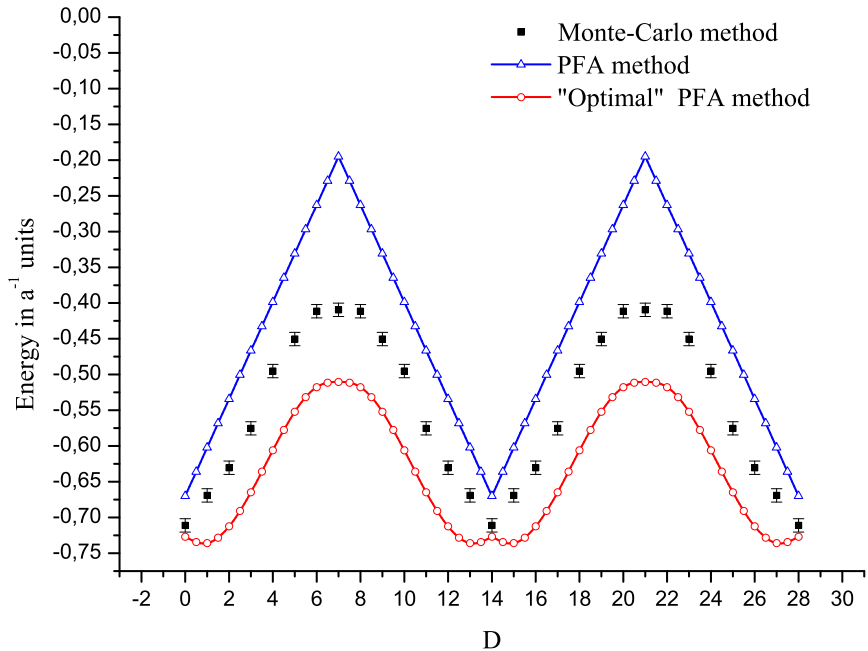} \end{center} \caption{Vacuum energy dependence on the lateral shift of the
 gratings. Comparison of our Monte-Carlo calculations with the results of PFA and ``Optimal''
 PFA methods.
 Lattice sizes: $42 a \times 42 a \times 42 a \times 22 a_z$. Deformation coefficient: ${a_z / a_{xyt} }={1 / 3}.$}
\end{figure}

We also compare our calculation with the results of approximate
methods (PFA and ``Optimal'' PFA \cite{opfa}). The PFA approach is
often considered as the initial approximation for such systems. In
the case of the two gratings interaction the PFA method yields the
``saw'' consisted of the straight lines. It is obvious that the
deviations from the PFA are rather large, especially when the
``teeth'' of one grating are opposite to the wells of the other
one. So a corrections to ordinary PFA should be considered. We
compare our results with so-called ``Optimal'' PFA. It was
proposed by Jaffe and Scardicchio in the paper \cite{opfa}. In
this approach the Casimir energy is given by:
$$
E=-{\pi^2 \over 720} \int_D d^3 \vec{r} {1 \over
(l_{12}(\vec{r}))^4}.
$$
Here $D$ is space region between the boundaries and
$l_{12}(\vec{r})$ is the shortest path that passes through the
point $\vec{r}$ from one boundary to another. Unfortunately in
this case of rectangular gratings ``Optimal'' PFA is not fully
adequate too (see Fig. ~5).

We also calculate the vacuum energy density that corresponds to
the gratings interaction. It is shown in Fig.~6 for different
lateral shifts for Monte-Carlo calculations and for ``Optimal''
PFA.
\begin{figure}[h]
 \epsfysize=160mm \epsfbox{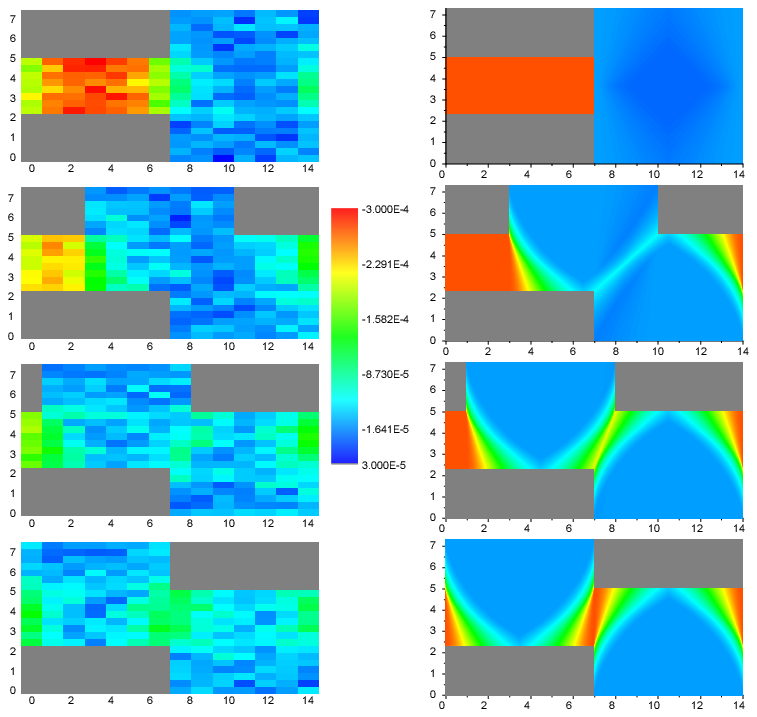} \caption{Vacuum
energy density distribution for different lateral shifts. Results of
Monte-Carlo calculations (left column) and ``Optimal'' PFA
calculations (right column).}
\end{figure}

The calculation of the energy density (see Fig.~6) allows us to
clear up the reasons why the PFA is not appropriate for this shape
of interacting surfaces. In the PFA approach, we take into account
only interaction of the parts of the surfaces that lie opposite to
each other. One can see in Fig.~6 that in the situation when the
``teeth'' are opposite to the ``wells'', the vacuum energy density
is concentrated in the space region between the corners of the
``teeth''. And it is just the very case when the PFA gives maximal
error. Hence we can conclude that the main reason for the errors
in the PFA is the neglecting of those specific Casimir forces that
appear between the ``corners'' of the gratings.

``Optimal'' PFA takes into account the ``corners'' interaction.
However its weak point in this geometry is the edge effect: energy
density in the region between parallel parts of the gratings
strongly depends on the size of this region (see the first column
in Fig. ~6), but in PFA and ``Optimal'' PFA (see the second column
in Fig. ~6) energy density in this region is constant.

Our method can also be considered as Monte-Carlo calculation of
some elements of the Green's function. So for abelian fields our
results will be close to the results obtained by the Euclidean
Green's function method \cite{mit1}.
%%%%%%%%%%%%%%%%%%%%%%%%%%%%%%%%%%%%%%%%%%%%%%%%%%%%%%%%%%%%%%%%%%%%%%%%%%%%%%%
\section*{Conclusions}

We introduced the new method for calculation of the Casimir forces
based on the Monte-Carlo simulations in noncompact QED on the
lattice. The method consists in two parts: the definition of the
boundary conditions for the field configurations; the definition
of the lattice variable that gives the vacuum energy density after
the averaging over these configurations.

The method is tested on the problem of the Casimir interaction of
two parallel conducting plates. Also this approach is used to
study the lateral Casimir forces between two rectangular gratings.
The Casimir energy and the vacuum energy density for the
interaction are obtained for different lateral shifts of the
gratings.

\section*{Acknowledgements}

The authors thank I. G. Pirozhenko and V. N. Marachevsky for the
substantial discussions. Computing facilities of the MSU
Supercomputer Center have been used for our calculations.

\end{document}